# Rotating spoke instabilities in a wall-less Hall thruster: Simulations.

K. Matyash[1], R. Schneider[1], S. Mazouffre[2], S. Tsikata[2] and L. Grimaud[2]

[1] Greifswald University, Greifswald, D-17487, Germany.
[2] ICARE, CNRS, 1C avenue de la Recherche Scientifique, 45071 Orléans, France.

**Abstract**

The low-frequency rotating plasma instability (spoke) in the ISCT200 thruster operating in the wall-less configuration was simulated with a 3 dimensional PIC MCC code. In the simulations an $m = 1$ spoke rotating with a velocity of 6.5 km/s in the $E$x$B$ direction was observed. The rotating electron density structure in the spoke is accompanied by a strongly depleted region of the neutral gas, which clearly shows that the spoke instability is of an ionization nature, similar to the axial breathing mode oscillations. In the simulation the electron cross-field transport through the spoke core was caused by diffusion in the high-frequency (4-10 MHz), short-scale (3 mm) electric field oscillations. These short-scale oscillations play a crucial role in the thruster discharge as over 70% of the electron current to the anode originates from the spoke core. The rest of the current originates from the spoke front where the electron cross-field transport toward the anode is due to the $E$x$B$ drift in the spoke macroscopic azimuthal electric field.

## 1. Introduction

The low-frequency rotating plasma instability, often called the "rotating spoke", is a phenomenon frequently observed in various $E$x$B$ cross-field discharges like Hall thrusters, magnetrons and plasma columns [1]. In Hall thrusters, despite decades of investigation, the origin, dynamics and nature of the spoke are still poorly understood. The earliest spoke study, carried out by Janes and Lowder [2], identified an azimuthal structure moving at tens of kilometers per second, whose presence was later attributed to the ionization type instability [3]. Our recent Particle-in-Cell with Monte Carlo collisions (PIC MCC) simulations of the cylindrical Hall thruster (CHT) [4] have also pointed out the ionization nature of the spoke instability. However, this hypothesis still needs to be validated both theoretically and experimentally.

In a Hall thruster, the spoke instability is of prime importance because it has been shown to be present under any operating conditions, with the mode number varying according to the thruster size and operating conditions [5].

The rotating spoke is of particular interest because it can conduct current, thereby participating in anomalous electron transport across the magnetic field [5-7]. Measurements performed in a CHT thruster have demonstrated that a large fraction of electron current towards the anode flows through the spoke [6,7]. Probe measurements of the time-evolution of the azimuthal electric field also revealed that the electric field generated by the spoke is responsible for axial electron current.

In this work we apply the 3D PIC MCC code STOIC [4,8,9] to investigate the properties of low-frequency rotating plasma instabilities in the discharge of the low-power ISCT200 Hall thruster [10,11] operating in the wall-less configuration [12-15]. The main objective was to characterize the dynamics of rotating plasma structures in the $E$x$B$ discharge of ISCT200-WL and to provide the insight into the physics of the rotating spoke instability by means of self-consistent PIC simulations.

## 2. ISCT200-WL Hall thruster

The Hall thruster simulated in this work is the ISCT200 thruster, where ISCT is an acronym for ICARE Small Customizable Thruster. The ISCT200 is a versatile 200 W-class Hall thruster using permanent magnets for generating the magnetic field instead of helical magnetizing coils [10,11]. The ISCT200 thruster can be operated in the standard (ST) and the wall-less (WL) configurations. The wall-less configuration is a recent and original approach to decrease plasma-wall interactions, hence prolonging the thruster lifetime [12,13]. The principle is to entirely shift the ionization and acceleration regions outside the cavity. This unconventional design is named a Wall-Less Hall Thruster, or WL-HT in short [12-16]. This configuration was in fact first explored during the 90s in Ukraine by Kapulkin et al. [17,18]. To shift both the ionization and the acceleration anode is a 1mm thick grid with 3 mm-diameter circular holes placed exactly at the channel exit. The anode transparency is 0.68, which allows a homogeneous propellant gas distribution outside the channel. A photograph of the ISCT200 Hall thruster in WL configuration with its gridded anode at the channel exit is displayed in Figure 2. The same figure shows the ISCT200-WL thruster firing with Xe at 1 mg/s and 200V discharge voltage in the NExET vacuum chamber. As can be seen from this image, the plasma discharge is detached from the anode - the signature of an efficient confinement.

A wall-less ion source provides an ideal platform for the study of cross-field discharge configurations with the probes and optical diagnostic tools. The access it provides to key regions of the plasma facilitates a thorough investigation of plasma instabilities and small scale turbulence for a better regions outside the thruster cavity, the anode is moved towards the channel exit. Schematics of the ISCT200 thruster both in the standard and the wall-less configurations are presented in Figure 1. In the wall-less configuration the understanding of the discharge physics and anomalous electron transport. Additionally, it allows these phenomena to be studied without the influence of wall processes such as secondary electron emission and sputtering. A wall-less Hall thruster is also an adequate configuration to make comparisons between experiments and computer simulations. Due to a simplified architecture, boundary conditions are easier to model, which permits more accurate and reliable numerical outcomes to be obtained. For all the aforementioned reasons, the wall-less configuration has been selected to investigate the physics of rotating plasma instabilities in this work.

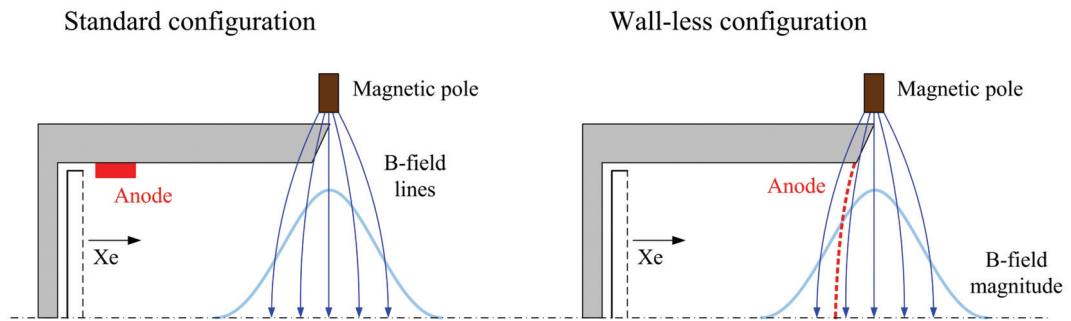

**Figure 1.** Schematic of a ISCT200 thruster in the standard (left) and wall-less (right) configurations.

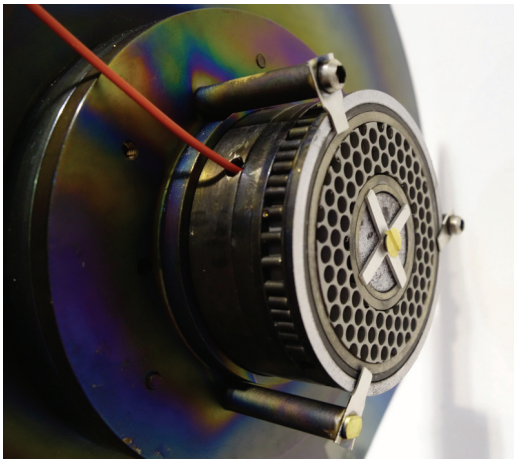
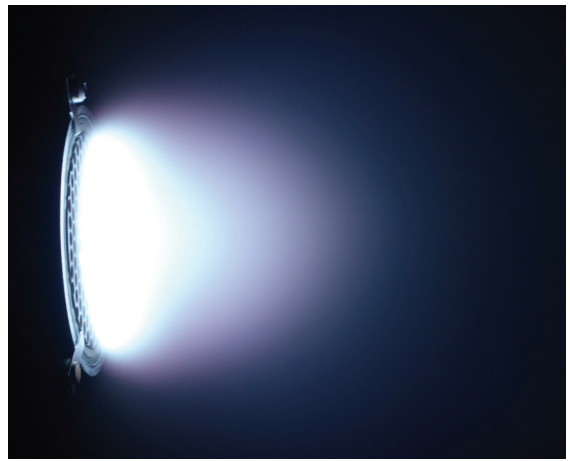

**Figure 2.** Picture of the ISCT200 Hall thruster in WL configuration with a gridded anode at the channel exit plane (left) and photograph of the thruster firing with Xe in the NExET test bench.

## 3. The model

In this work, the self-consistent 3D-3V Particle-in-Cell simulation code STOIC (electrostatic Optimized particle In Cell) [4,8,9] has been applied to simulate the rotating spoke instability in the ISCT200 thruster operating in the wall-less configuration [13-15]. The simulation includes electrons, $Xe^+$ ions and neutral Xenon atoms. All relevant collisional processes are included in the model via Monte Carlo collision algorithms: Coulomb collisions between charged particles; electron-neutral elastic, ionization and excitation collisions; ion-neutral momentum transfer and charge exchange collisions and neutral-neutral elastic collisions. The dynamics of the background neutral gas is self-consistently resolved with direct simulation Monte Carlo (DSMC). The model is full-3D - 3 spatial and 3 velocity components are resolved. It utilizes an equidistant Cartesian grid which explicitly assures momentum conservation and zero self forces.

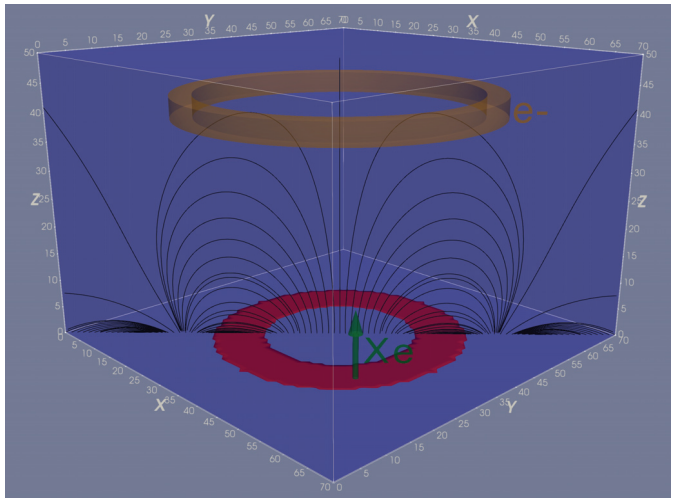

**Figure 3.** Computational domain with magnetic field topology and particle sources.

The computational domain represents a cuboid with length $Z_{max} = 50$ mm and sides $X_{max} = Y_{max} = 70$ mm. The Z axis is directed along the thruster symmetry axis. The sketch of the computational domain together with magnetic field topology and particle sources is shown in Fig. 3. The plane $Z = 0$ corresponds to the channel exit, where the anode (red) is placed. The ring anode is mapped with the Cartesian mesh. All boundaries of the computational domain are assumed to be metallic. All metal elements in the simulation, except for the anode are at ground potential (blue). At the anode, a voltage of $U_a = 225$ V is applied. The neutrals are injected into the system through the anode with the mass flow rate $\dot{m} = 1$ mg/s and $T_n = 400$ K. Electrons with a Maxwellian distribution and a temperature $T_e = 2$ eV are introduced into the system in the source region $42\text{ mm} < Z < 46\text{ mm}$, $26\text{ mm} < R < 30\text{ mm}$ with uniform density and the constant current $I_c = 0.25$ A, simulating the thruster cathode. All surfaces in the simulation are assumed to be absorbing for electrons and ions. No secondary electron emission is included in the simulation. The neutrals are lost on all surfaces, except for the anode plane, where they are re-launched with a Maxwellian distribution with the temperature $T_n = 400$ K.

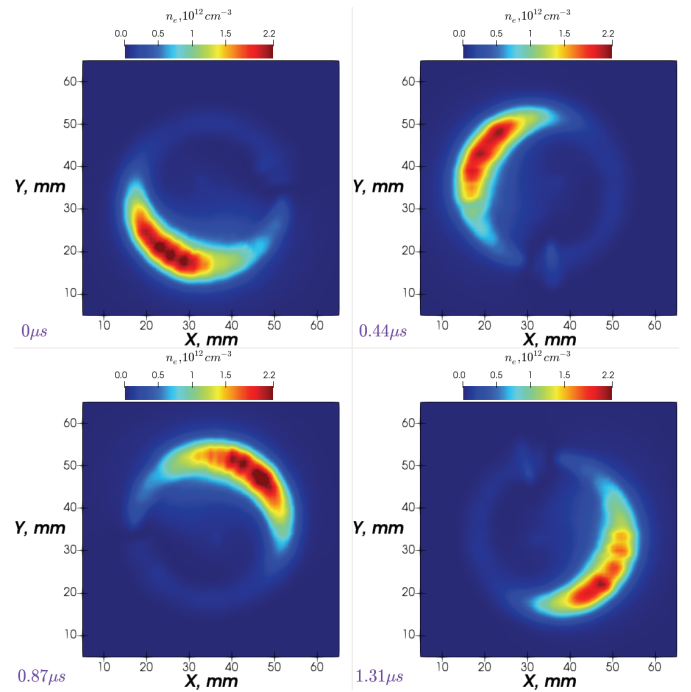

**Figure 4.** Evolution of the plasma density 3 mm above the anode during the spoke cycle. The spoke $m = 1$ rotates in the $ExB$ direction with a velocity of 6.5 km/s.

To reduce the computational time the size of the system is scaled down by a factor of 10. In order to preserve the ratio of the particles' mean free paths and the gyroradii to the system length, the collision cross-sections and the magnetic field are increased by the same factor of 10.

An equidistant computational grid 70x70x50 was used in the simulation. The total number of computational particles in the simulation was about $1.6 \cdot 10^8$. The cell size $\Delta x = \Delta y = \Delta z = 10^{-1}$ mm in the simulation was chosen to ensure that it is smaller than the smallest Debye length in the system. The time step was set to $\Delta t = 5.6 \cdot 10^{-12}$ s in order to resolve the electron plasma and cyclotron frequencies. The simulation was carried on a 16-processor Intel Xeon workstation, with a run duration of about 50 days. About $1.8 \cdot 10^7$ time steps were performed, corresponding to a simulated time of 100 μs.

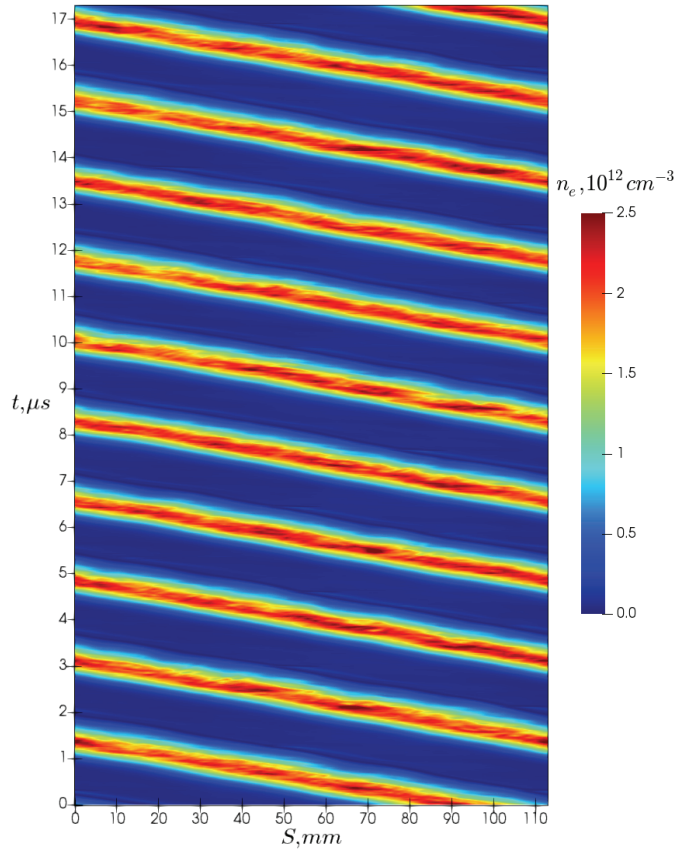

**Figure 5.** Evolution of the plasma density along the spoke median line (circle of radius 18 mm, 3 mm above the anode) during 10 spoke cycles. The spoke $m = 1$ rotates in the *ExB* direction with a velocity of 6.5 km/s.

## 4. Simulation results

In the simulation, after a start up transient of about 5 μs, the rotating spoke instability develops in the ionization region close to the anode. In Figure 4 the evolution of electron density *XY* cross-section 3 mm above the anode during the spoke cycle is shown. The $m = 1$ mode spoke is rotating in the *ExB* direction (clockwise) with the average velocity 6.5 km/s, which corresponds to the rotation frequency of 57 kHz in the real (unscaled) geometry. The spoke rotation velocity in the simulation is about a factor of 3 higher than observed in the experiment [14,15]. Such discrepancy with the experimental results may be caused by the proximity of the domain boundaries to the discharge plasma in the simulations and by the applied geometrical scaling. After its development the spoke is present until the end of the simulated time of 100 μs and demonstrates a very regular pattern. The spoke rotation period fluctuation level throughout the simulation is below 5%. In Figure 5 the evolution of the plasma density along the spoke median line (the circle of radius 18 mm, 3 mm above the anode) during 10 spoke rotation cycles is shown. In total 55 full spoke cycles occur during the simulated time.

In the previous simulations in addition to the $m=1$ spoke rotating in the *ExB* direction, we also observed $m=1$ and $m=2$ spokes rotating in the counter-*ExB* direction [9]. The spoke rotation direction and the mode number were dependent on the applied anode voltage and neutral gas flow from the anode source.

In Figure 6 the evolution of the electron and neutral density, ionization rate and plasma potential along the spoke median line during the spoke cycle is presented. The spoke moves from right to left, which corresponds to clockwise rotation. Higher plasma density in the spoke results in the local elevation of the plasma potential, which reaches up to 260 V, higher than the anode voltage $U_a = 225$ V. The rotating electron density structure is accompanied by the strongly (up to 75%) depleted region of the neutral gas, which clearly shows that the spoke instability is of the ionization type. As the largest part of the neutrals are locally consumed by electron-impact ionization, the ionization front moves further toward the region with higher neutral density. The void in the neutral gas, which is left behind, then is gradually refilled by the gas flux from the anode source, such that the neutral density is restored to the previous level before the ionization front makes the full turn, so that the cycle can recur. This process looks rather similar to the breathing mode oscillations [19, 20], where the neutral gas front moves back and forth in the axial direction due to predator-prey-like dynamics between the electrons and the neutrals. But in contrast to the breathing mode, the spoke instability does not manifest itself in the discharge current, as the plasma density averaged over the discharge transverse cross-section stays constant as the spoke rotates.

Another important feature visible in Figure 6 is that in addition to the macroscopic spoke structure, there are short-scale (2-4 mm) oscillations present in the electron density, plasma potential and the ionization rate. These short-scale oscillations have the frequencies in the in 4-10 MHz range [9] and likely are responsible for the electron transport through the spoke core.

In fact, recent measurements we have made using coherent Thomson scattering on a wall-less version of a 1.5 kW thruster indeed show the coexistence of oscillations in the same high-frequency range (associated with the electron cyclotron drift instability [21]) with spokes. These observations will be discussed in detail in future work. The small-scale turbulence inside the rotating spoke was also observed in the CHT thruster experiment [22].

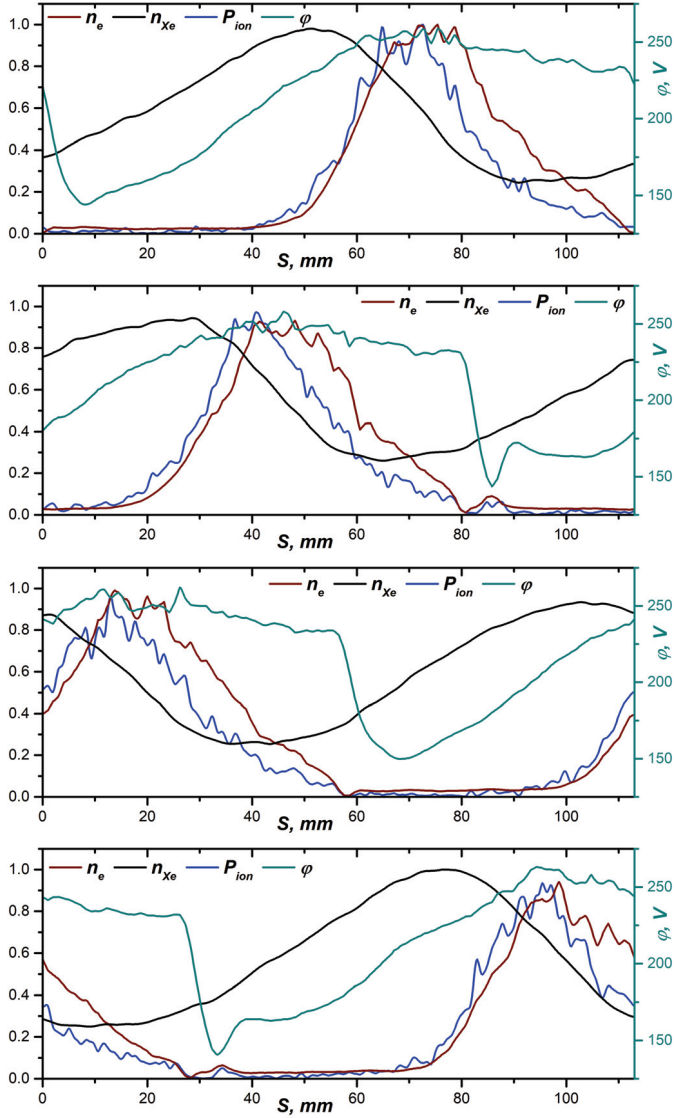

**Figure 6**. Electron and neutral density, ionization rate and plasma potential along the spoke median line (circle of radius 18 mm, 3 mm above the anode) during the spoke cycle. From top to bottom: $t=0\mu s$, $t=0.44\mu s$, $t=0.87\mu s$, $t=1.31\mu s$. $n_e$ is normalized by $2.3\cdot 10^{12}$ cm$^{-3}$, $n_{Xe}$ – by $1.2\cdot 10^{13}$ cm$^{-3}$, $P_{ion}$ – by $2.7\cdot 10^{18}$ cm$^{-3}$s$^{-1}$.

In Figure 7 the axial and azimuthal electric field along the spoke median line are plotted. In the spoke core ($61\,\text{mm} < X < 103\,\text{mm}$) short-scale, high-frequency (HF) oscillations with the electric field amplitude about 100 V/cm (both axial and azimuthal) are present. The mean electric field in the spoke core is much lower, below 5 V/cm (both axial and azimuthal), thus the regular azimuthal $E$x$B$ drift of electrons inside the spoke bulk is impossible and the electron dynamics is determined by the short-scale HF oscillations. The axial electric field, averaged over the spoke cycle, is about 60 V/cm, which corresponds to an $E$x$B$ drift velocity of 188 km/s. This is about 30 times higher than the spoke velocity of 6.5 km/s.

The regular azimuthal $E$x$B$ drift is possible only at the spoke front ($3\,\text{mm} < X < 61\,\text{mm}$) and the spoke rear ($103\,\text{mm} < X < 113\,\text{mm} \cup 0\,\text{mm} < X < 3\,\text{mm}$) where the axial electric field does not change its sign. In the spoke front the axial electric field grows from 50 V/cm to 200 V/cm so that the azimuthal electron $E$x$B$ drift is accelerated by factor 4 – from 157 km/s to 630 km/s. The azimuthal electric field for most of the spoke front ($12\,\text{mm} < X < 61\,\text{mm}$) is directed counter-clock-wise with the average value about 20 V/cm, which means that there exists an axial electron $E$x$B$ drift directed toward the anode with average velocity of about 63 km/s, which contributes to the anode current. These two effects: rapid acceleration of the azimuthal electron drift and the loss of the electrons at the anode, contribute to a decrease of the electron density at the spoke front, preventing the spoke front from flattening.

In the border region between the spoke rear and the front ($0\,\text{mm} < X < 12\,\text{mm}$), where the plasma potential fall of about 100 V occurs, the azimuthal electric field is directed clock-wise, reaching up to 178 V/cm, so the electron axial $E$x$B$ drift here has maximum value of 556 km/s and is directed up from the anode.

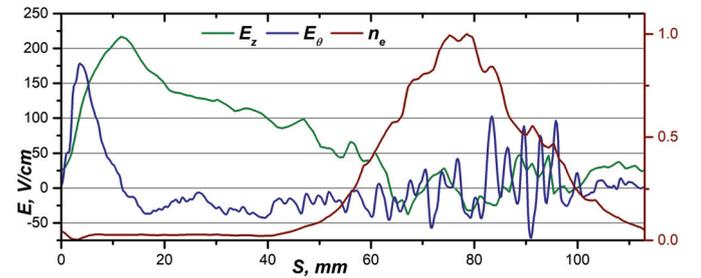

**Figure 7**. Axial and azimuthal electric field along the spoke median line. Normalized electron density is plotted for the reference.

In Figure 8 the isosurfaces of the electron density and the electric potential directly over the anode are presented. Here one can see that spoke has a complicated 3-dimensional structure. The asymmetry in the electron density due to the spoke leads to the asymmetry in plasma potential. As the electric field is normal to the potential isosurfaces and the magnetic field lines mostly follow them, the electron $E$x$B$ drift orbits should stick to surfaces of the constant potential. Thus asymmetry in potential distribution should lead to asymmetry in the electron drift motion in the spoke region. As one can see, the isosurface of the $\varphi = 235\,\text{V}$ in the spoke core is strongly distorted due short-scale oscillations, thus, again, one should not expect the regular azimuthal electron *ExB* drift in this region.

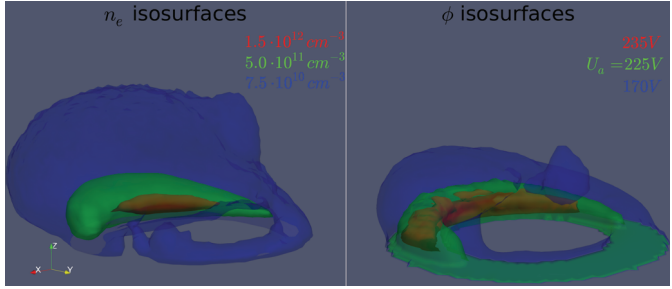

**Figure 8**. The isosurfaces of the electron density and the plasma potential over the anode.

Further insight into the electron transport through the spoke can be gained by investigating the electron trajectories. In Figures 9-12 the typical trajectories of the test electrons in the spoke region during the 84 ns are plotted. For orientation purpose the potential isosurface of $\varphi = 200$ V is plotted as well. As can be seen in Figures 9-10, in the spoke core electron motion has a clear diffusive pattern, with both the particle position and the energy changing randomly (the electron energy can be estimated from the electron gyroradius). The test electrons do not collide with other particles, and their motion is completely defined by background electric and magnetic fields. Thus, indeed, strong high-frequency, short-scale electric field oscillations in the spoke core disrupt the regular *ExB* drift of the electrons, and lead to diffusion-like motion in this region. As a result of this diffusive motion some of the electrons reach the anode, contributing to the anode current. One can therefore claim that in the simulation the electron current through the core of the spoke is due to diffusion in the HF electric field oscillations. These oscillations play a crucial role in the thruster discharge since over 70% of the electron current to the anode originates from the spoke core.

Analyzing the electron motion at the spoke front (see Figure 11) one can distinguish the clear *ExB* drift pattern in the spoke macroscopic electric field superposed with the oscillations between magnetic mirrors along the magnetic field lines. The *ExB* drift in the axial direction due to the spoke azimuthal electric field, as expected, is directed toward the anode. In Figure 11 it can be seen how, as a result of this drift motion, electrons reach the anode, landing support to the idea that the electron current through the spoke front is due to the *ExB* drift in the spoke macroscopic azimuthal field.

In Figure 12 the motion of the test electron at the spoke rear can be observed. It is clearly an *ExB* drift combined with the bouncing between magnetic mirrors along the B-field lines. As expected, in the border region between the spoke rear and the front, the electron *ExB* drift in the axial direction is directed up from the anode. The electron drift orbit in Figure 12 closely follows the equipotential surface, repeating its asymmetry, which is caused by the presence of the spoke.

## 5. Conclusion

A 3D PIC MCC simulation has been used to study the rotating spoke in the ISCT200-WL thruster. The $m = 1$ spoke rotating with a velocity of 6.5 km/s in the *ExB* direction was observed. The rotating electron density structure in the spoke is accompanied by a strongly depleted (up to 75%) region of the neutral gas, which provides evidence that the spoke instability is an ionization-driven phenomenon, similar to the axial breathing mode oscillations.

Strong high-frequency electric field oscillations, with frequency in the range of 4-10 MHz, length scale of about 3 mm and amplitudes of the order of 100 V/cm were observed inside the spoke.

These electric field oscillations have characteristics very similar to the electron cyclotron drift instability (frequency, length scale) observed experimentally in recent years in both thrusters and planar magnetrons [23,24], and are further evidence for its likely role in electron transport. In the simulation the electron cross-field transport through the spoke core was caused by diffusion in the HF, short-scale electric field oscillations, a result which appears to account for the observation that over 70% of the electron current to the anode originates from the spoke core. The rest of the current originates from the spoke front, where the electron cross-field transport toward the anode is due to the *ExB* drift in the spoke macroscopic azimuthal electric field.

The electron drift orbits at the spoke region are axially asymmetric, consistent with the asymmetry of the electric potential caused by the presence of the spoke.

Further joint efforts of simulation and experiment are ongoing for clarification of the phenomena underlying spoke formation and dynamics, and in particular, the link to the HF oscillations responsible for the electron current in the spoke core.

**Acknowledgements**

This work was supported by European Office of Aerospace Research and Development (EOARD) grant FA9550-15-1-0281.

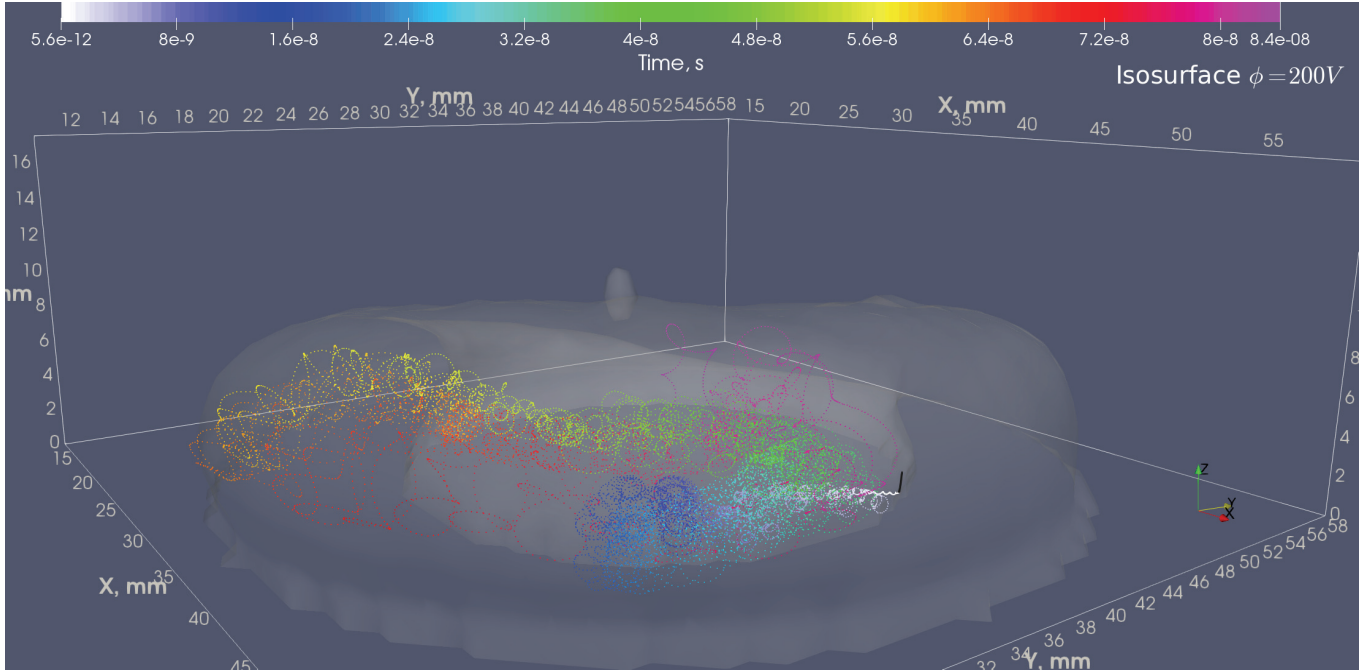

**Figure 9.** Trajectory of the test electron in the spoke core during 84 ns. The potential isosurface $\varphi = 200$ V is plotted for the reference.

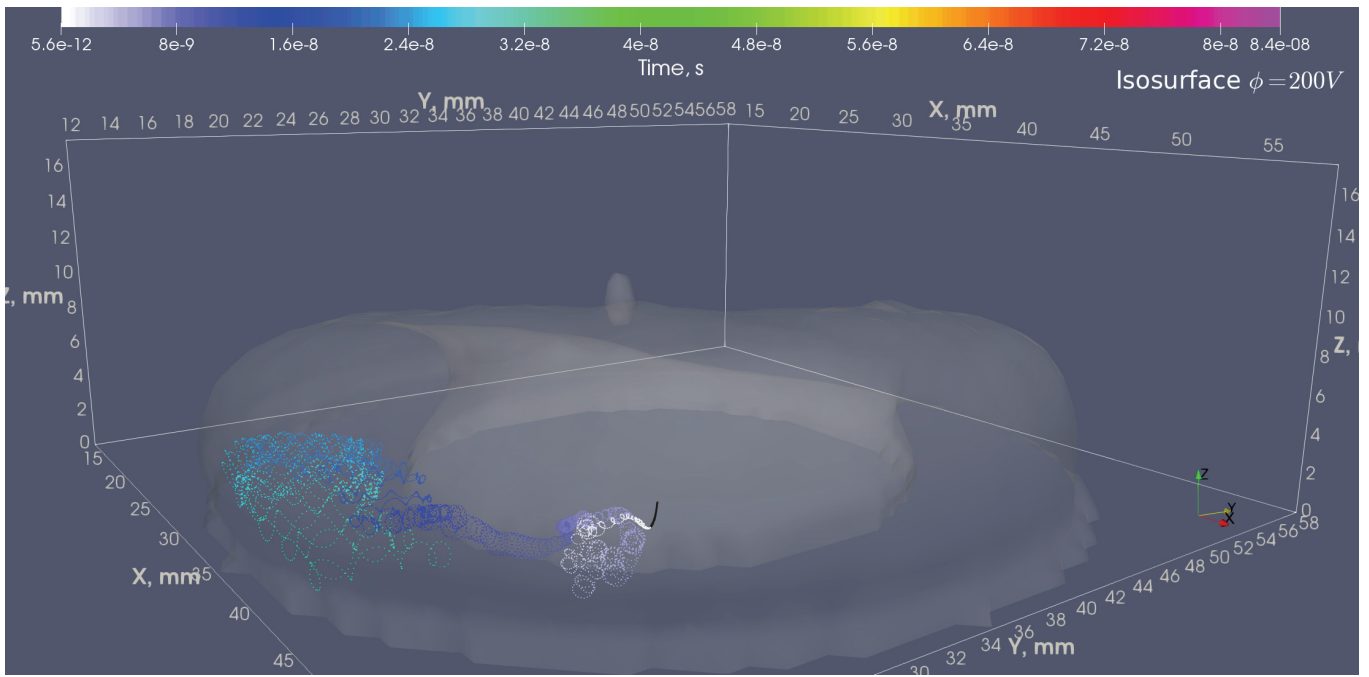

**Figure 10.** Trajectory of the test electron in the spoke core during 84 ns. The potential isosurface $\varphi = 200$ V is plotted for the reference.

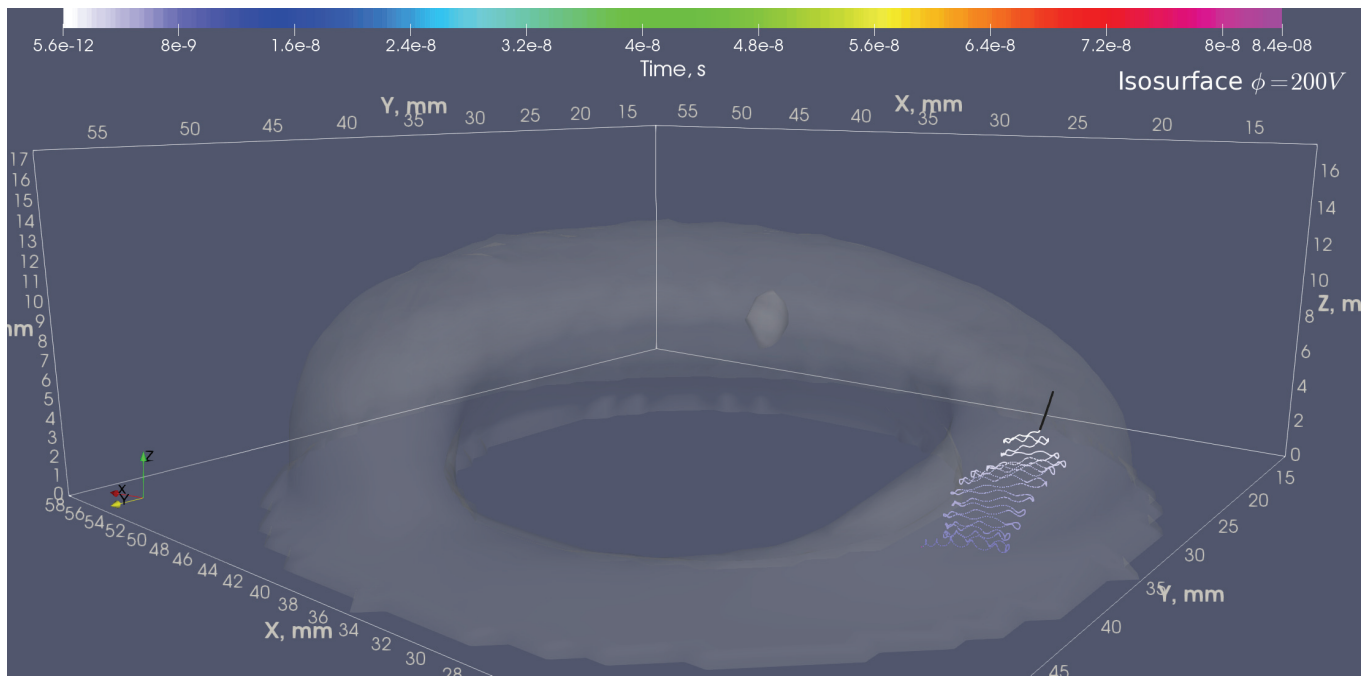

**Figure 11.** Trajectory of the test electron at the spoke front during 84 ns. The potential isosurface $\varphi = 200$ V is plotted for the reference.

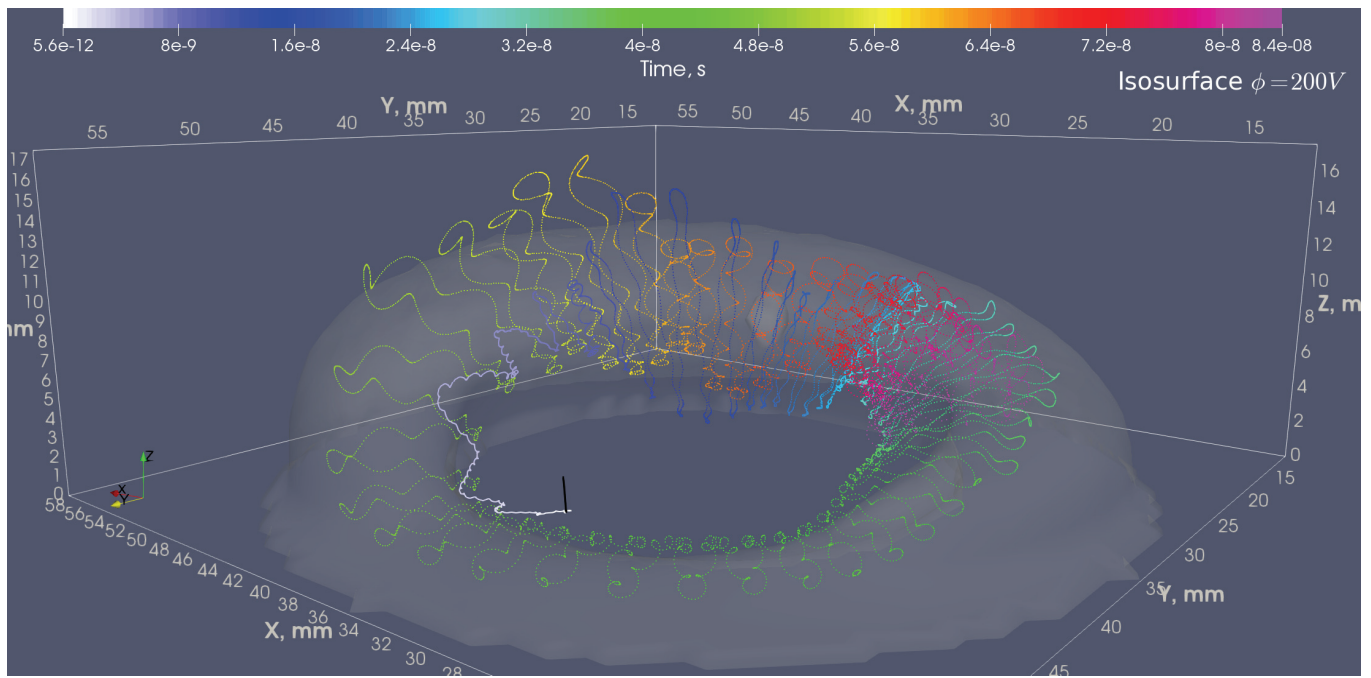

**Figure 12.** Trajectory of the test electron at the spoke rear during 84 ns. The potential isosurface $\varphi = 200$ V is plotted for the reference.